\documentclass{article}

\usepackage{arxiv}

\usepackage[utf8]{inputenc} 
\usepackage[T1]{fontenc}    
\usepackage{hyperref}       
\usepackage{url}            
\usepackage{booktabs}       
\usepackage{amsfonts}       
\usepackage{nicefrac}       
\usepackage{microtype}      
\usepackage{lipsum}

\usepackage{graphicx}

\usepackage{amsmath}

\title{Anonymous Authentication using Attribute-based Encryption}

\author{
  Nouha Oualha\\
  Université Paris-Saclay, CEA, List, F-91120, Palaiseau, France\\
  \texttt{nouha.oualha@cea.fr} \\
}

\begin{document}
\maketitle

\begin{abstract}
In today's digital age, personal data is constantly at risk of compromise. Attribute-Based Encryption (ABE) has emerged as a promising approach to privacy-preserving data protection. This paper proposes an anonymous authentication mechanism based on ABE, which allows users to authenticate without revealing their identity. The mechanism adds a privacy-preserving layer by enabling authorization based solely on user attributes. The proposed approach is implemented using OpenID Connect, demonstrating its feasibility in real-world systems.
\end{abstract}


\section{Introduction}
As digital systems become increasingly pervasive, concerns surrounding privacy and security continue to grow. To address these concerns, a variety of cryptographic primitives have been proposed to protect sensitive data. Among them, Attribute-Based Encryption (ABE) stands out for offering not only public-key encryption but also fine-grained access control. When applied to user authentication, ABE allows decisions to be made based solely on the attributes presented by the user—rather than its identity—thus providing a cloak of anonymity while still allowing the system to verify the user's permissions. 

Several ABE-based anonymous authentication schemes have been proposed in the literature (e.g., \cite{TS103532}, \cite{Bartolomeo2023}, \cite{Portnoi2015}). However, many of these approaches fall short in terms of cryptographic robustness. In particular, they fail to ensure forward secrecy, a critical property that protects past sessions even if long-term keys are later compromised. This weakness exposes systems to quantum-capable adversaries and long-term surveillance risks.

To address these limitations, this paper introduces a novel ABE-based anonymous authentication mechanism. The proposed scheme achieves perfect forward secrecy through the use of ephemeral session keys, and it involves only a single message exchange for authentication.

\textbf{Paper Outline.} The remainder of this paper is structured as follows. Section 2 reviews related work. Section 3 provides background on ABE and its use in anonymous authentication. Section 4 describes the proposed scheme as a simple protocol based on ABE. Section 5 analyses the scheme’s compliance with privacy and security requirements. Section 6 discusses the implementation and integration of the scheme with OpenID Connect. Section 7 identifies open challenges and future directions. Finally, section 8 concludes the paper.

\section{Related work}
Anonymous authentication, while seemingly paradoxical, refers to a process in which a user proves eligibility to access a service without revealing its actual identity \cite{Lindell2011}. This paper focuses specifically on user anonymity, rather than broader privacy features such as unlinkability or untraceability. The user keeps its identity private, but its corresponding actions can be comprised, inferred, or linked e.g., through the IP address or some other information.

Traditional authentication approaches mainly include password authentication, cryptography-based authentication, and biometrics. To provide anonymity, approaches can obviously use pseudonyms instead of real identities or leverage base cryptographic schemes such as group signatures, ring signatures, attribute-based signatures (ABS) \cite{cryptoeprint:2010/595} and attribute-based encryption (ABE) (e.g., \cite{Bethencourt2007}, \cite{Waters2011}, \cite{Agrawal2017}, \cite{Lewko2011}, \cite{Chase}, \cite{Rouselakis2015}).

Several pseudonymous-based authentication schemes have been proposed in the context of vehicular networks and the IEEE 1609.2 standard, where vehicles need to frequently change their pseudonyms to avoid being tracked by adversaries \cite{Shahrouz2024}. Other anonymous authentication schemes have leveraged group signature schemes that preserve the privacy of users by allowing valid group members to sign anonymously a message on behalf of the whole group, but require selecting a group leader as a central authority, which makes these schemes prone to the escrow problem. On the other hand, a ring signature scheme does not require a group leader or a fixed group. Rather, all ring members are chosen in an ad hoc manner by a signer and have equal status \cite{Lindell2011}. Ring signatures usually do not provide traceability. Instead of proving group membership, ABS and ABE schemes enable users to convince others that they satisfy a specified policy or predicate \cite{Shahrouz2024}. Moreover, variants of these schemes like functional credentials \cite{Deuber2018} have been crafted to allow users to anonymously prove possession of a set of attributes that fulfils a certain policy or predicate.

This paper is interested in anonymous authentication leveraging attribute-based encryption (ABE) schemes. ABE can be applied to user authentication and implicitly enforce access control to resources and data. A simple challenge-response authentication mechanism based on ABE has been defined in \cite{TS103532} and \cite{Bartolomeo2023}. During the authentication process, a challenge token is encrypted using ABE. If the user possesses the keys containing the right set of attributes that satisfy the access policy, the user can send the correct response. Similarly in \cite{Portnoi2015}, the authentication approach proposed to encrypt with ABE a nonce exchanged during the authentication handshake. If it is able decrypt the nonce using the right set of attributes, the user is then able to generate the session  key. The approach provides user authentication using ABE-encrypted broadcast messages and location as an additional factor. Compared to these existing approaches that employ ABE to hold a basic anonymity feature, this paper endeavours to provide anonymity along with critical security properties not yet addressed, like forward secrecy.

\section{Preliminaries}
The section provides a background overview of ABE and ABE-based anonymous authentication.

\subsection{Attribute-based encryption}
The attribute-based encryption (ABE) concept was first introduced by Sahai and Waters in \cite{Sahai2005} as a fuzzing identity-based encryption scheme, and has been further developed in the literature (e.g., \cite{Bethencourt2007}, \cite{Waters2011}, \cite{Agrawal2017}, \cite{Lewko2011}, \cite{Chase}, \cite{Rouselakis2015}) to allow for fine-grained access control on encrypted data based on attributes. In ABE, identity is represented as a set of descriptive attributes, and these attributes are used to decrypt the ciphertext. 

\textbf{Attribute-based encryption types.} ABE comes in two different forms. The Ciphertext-Policy Attribute-based Encryption (CP-ABE) scheme associates a set of attributes to the user private keys and an access policy to the ciphertext, such that only the users possessing the attributes that satisfy the access policy associated with the ciphertext can decrypt it. On the other hand, in Key-Policy Attribute-based Encryption (KP-ABE) schemes, attribute sets are used to annotate ciphertexts while the user private keys are associated with the access policy that specifies which ciphertexts the user is entitled to decrypt. 

\textbf{Monotone Span Programs.} Access policies can be represented using Boolean formulae (e.g., AND and OR gates), but today, the majority of ABE schemes have expressed access policies using a more general representation, called Monotone Span Programs (MSPs). The access policy is expressed using an MSP encoding over the attributes in the policy. The authors in \cite{Lewko2011} (appendix G) showed a simple method to convert a Boolean formula with AND and OR gates into an MSP matrix with 0, ±1 values. The encoding of a policy from a Boolean representation $P$ to MSP generates a pair $(M,label)$ where $M$ is a matrix with integer entries and $label$ is a mapping that labels the lines of $M$ with the attributes occurring in $P$.

\textbf{Hybrid attribute-based encryption.} In public key encryption, it is common practice to rely on a hybrid cryptosystem combining the public encryption scheme with a symmetric encryption scheme. During this process, a random symmetric key is generated and encrypted using a Key Encapsulation Mechanism (KEM) built on the public key encryption scheme. The generated symmetric key is used to encrypt the actual data using a symmetric encryption scheme (e.g., AES). ABE schemes are not excluded from this rule. In practice, an attribute-based key encapsulation mechanism is employed to encrypt a randomly generated symmetric key that is used to encrypt data, while maintaining ABE benefits of fine-grained access control.

\textbf{Attribute-based key-encapsulation mechanism.} An Attribute-based Key Encapsulation Mechanism (ABKEM) consists of a hybrid encryption that aims to protect a symmetric key using attribute-based encryption. As specified in \cite{TS103532}, ABKEM comprises four algorithms. For instance, the algorithms of an ABKEM based on the Waters’ scheme \cite{Waters2011}, which is a CP-ABE variant, are presented in the following:
\begin{itemize}
\item \textit{$Setup(k) \rightarrow (params,mpk,msk)$:} The setup operation takes as input a security parameter $k$ and establishes the set of base groups $G_{1}$, $G_{2}$, $G_{T}$ and a pairing $e:G_{1} \times G_{2}\rightarrow G_{T}$. The base groups are established together with a generator $g_{1}$ of $G_{1}$ and a generator $g_{2}$ of $G_{2}$. Additionally, the operation defines a hash function $H$ that hashes strings into elements of $G_{1}$. The system parameters are defined as follows:
\begin{equation}
params=(k,p,G_{1},G_{2},G_{T},e,g_{1},g_{2},H)                                 
\end{equation}

The operation generates uniformly random variables $a$,$b$ in $\mathbb{Z}_{p}$ and sets the master public key as: 
\begin{equation}
mpk=(mpk_{1}, mpk_{2}) = (g_{1}^{b}, e(g_{1}, g_{2})^{a})                                \end{equation}
and the master secret key as:
\begin{equation}
msk=g_{1}^{a}
\end{equation}

The operation returns the system parameters $params$, the master public key $mpk$, and the master secret key $msk$.

\item \textit{$KeyGen(params,mpk,msk,S) \rightarrow sk_S$:} The secret-key generation operation takes as input the system parameters $params$, the master public key $mpk$, the master secret key $msk$, and a set of attributes $S=(s_{1},s_{2},\ldots,s_{t})$. The operations selects $r \in \mathbb{Z}_{p}$ uniformly at random and computes equation (4).
\begin{equation}
x_{1} = msk \cdot (mpk_{1})^{r} = g_{1}^{(a + br)} ,\ x_{2} = g_{2}^{r}
\end{equation}

For $i \in {1,\ldots,t}$, the operation computes equation (5).
\begin{equation}
sk_{i} = H(s_{i})^{r}
\end{equation}
The operation returns the secret key $sk_{S}=(x_{1},x_{2},sk_{1},\ldots,sk_{t})$.

\item \textit{$KeyEncap(params,mpk,MSP,R) \rightarrow (K,C_P)$:} The symmetric-key encapsulation operation takes as input the system parameters $params$, the master public key $mpk$, MSP encoding $(M_{P},label_{P})$, and a random integer $R$ serving as a seed for the pseudo-random number generation. Let $n$ be the number of lines of $M_{P}$ and $m$ the number of its columns. The operation pseudo-randomly generates $s,v_{2},\ldots,v_{m}$ in $\mathbb{Z}_{p}$ and computes equation (6).
\begin{equation}
(\mu_{1}, \ldots, \mu_{n})= M_{P} \cdot (s, v_{2}, \ldots, v_{m}) \mod p 
\end{equation}
The operation computes equation (7).
\begin{equation}
z = g_{2}^{s} ,\ K = mpk_{2}^{s} = e(g_{1} ,\ g_{2})^{as}
\end{equation}
For $i \in {1,\ldots,n}$, the operation pseudo-randomly generates $r_{i}$ in $\mathbb{Z}_{p}$ and computes $c_{i}=(c_{i,1},c_{i,2})$ defined in equation (8).
\begin{equation}
c_{i,1} = mpk_{1}^{\mu_{i}} \cdot H(label_{P}[i])^{-r_{i}} ,\ c_{i,2} = g_{2}^{r_{i}}
\end{equation}
The operation computes equation (9).
\begin{equation}
C_{P}=(z,c_{1},\ldots,c_{n})
\end{equation}
Finally, it returns $(K,C_{P})$.

\item \textit{$KeyDecap(params,MSP,C_P,S,sk_{S}) \rightarrow (K or \bot)$:} The symmetric-key decapsulation operation takes as input the system parameters $params$, an MSP encoding $(M_{P},label_{P})$, an encapsulation $C_{P}$,  a set of attributes $S$, and a secret decryption key $sk_{S}$. The decoding of MSP returns either the invalidity flag $\bot$ if the attributes do not satisfy the policy or a set of indices $I$ and a list of non-zero integers $d_{I}$ such that equation (10) holds.
\begin{equation}
\sum_{i \in I}d_{i}M[i]=(1,0,\ldots,0)                                          
\end{equation}
The operation computes equation (11).
\begin{equation}
w = \prod_{i\in I}c_{i,1}^{d_{i}} ,\ K = \frac{e(x_{1},z)}{e(w,x_{2} )\cdot \prod_{i \in I} e(sk_{pos(i)},c_{i,2}^{d_{i}})}
\end{equation}
where $pos(i)$ is the index of $label_{P}[i]$ in $S$. Finally, the operation returns $K$.
\end{itemize}

\section{Anonymous authentication: proposed construction}
This section first defines the requirements that should be met by the proposed approach with regard to security and privacy. Then, the section delves into the description of the employed \textbf{ABKEM} scheme and the authentication protocol.

\subsection{Security and privacy requirements}
The anonymous authentication scheme should meet the following security requirements:
\begin{itemize}
\item \textit{Secure authentication:} An unauthorised user should not be able to fool the server into granting it access.
\item \textit{Mutual authentication:} The two parties authenticate each other at the same time before establishing a connection.
\item \textit{Forward secrecy:} The generated session keys should not be compromised, even if "forward" long-term key corruption occurs (i.e., ABE keys are compromised).
\end{itemize}

The proposed scheme needs to be resilient against the following attacks:
\begin{itemize}
\item \textit{Replay attack:} The adversary maliciously or fraudulently repeat or delay a valid message in order to gain unauthorised access.
\item \textit{Man-in-the-middle attack:} The adversary positions itself in between the two parties to intercept and alter messages travelling between them.
\end{itemize}

The proposed anonymous authentication scheme needs also to fulfil the following privacy requirement:
\begin{itemize}
\item \textit{Anonymity:} The server should not be able to know the identity of the user.
\end{itemize}

\subsection{Hardness assumptions}
Any typical cryptographic scheme relies on hard computational problems to protect secret messages. The following defines two problems assumed to be hard in this paper:
\begin{itemize}
\item \textit{(decisional) Diffie-Hellman problem (DH):} For a chosen a group $G$ of prime order $p$ according to the security parameter, whose generator is $g$, let $a,b \in \mathbb{Z}_{p}$ be chosen at random. Given $g^{a}$, $g^{b}$ the adversary must distinguish $g^{ab}$ from a random element $T$ in $G$.
\item \textit{(decisional) Bilinear Diffie-Hellman problem (BDH):} For a chosen group $G$ of prime order $p$ according to the security parameter, whose generator is $g$ and a bilinear map $e: G \times G \rightarrow G_{T}$, let $a,b,c \in \mathbb{Z}_{p}$ be chosen at random. Given $g^{a}$, $g^{b}$, $g^{c}$ the adversary must distinguish $e(g,g)^{abc}$ from a random element $T$ in $G_{T}$.
\end{itemize}

\subsection{Proposed CP-ABKEM}
The proposed anonymous authentication based on attribute-based encryption relies on a \textbf{CP-ABKEM} construction (e.g., \cite{TS103532}, \cite{Bethencourt2007}, \cite{Waters2011}, \cite{Agrawal2017}, \cite{Lewko2011}). In this paper, the proposed approach is described using Waters scheme \cite{Waters2011} (refer to section 3.1) as example.

The proposed approach introduces a slight change in one algorithm in the \textbf{CP-ABKEM}. In the new \textbf{CP-ABKEM}, called \textbf{CP-ABKEM*} to highlight the change, the modification concerns the symmetric-key encapsulation operation that returns additionally the generated random secret s. The proposed \textbf{CP-ABKEM*} thus amended consists of the following algorithms:
\begin{itemize}
\item \textit{$Setup(k) \rightarrow (params,mpk,msk)$:} same algorithm as specified in section 3.1.
\item \textit{$KeyGen(params,mpk,msk,S) \rightarrow sk_S$:} same algorithm as specified in section 3.1.
\item \textit{$KeyEncap^{*}(params,mpk,MSP,R) \rightarrow (K,C_P,s)$:} In addition to the original operation (as specified in section 3.1, the random value s is kept and returned at the end of the operation.
\item \textit{$KeyDecap(params,MSP,C_P,S,sk_S ) \rightarrow (K or \bot)$:} same algorithm as specified in section 3.1.
\end{itemize}

\subsection{Proposed anonymous authentication}
\begin{figure}[ht!]
  \centering
  \includegraphics[width=0.5\columnwidth]{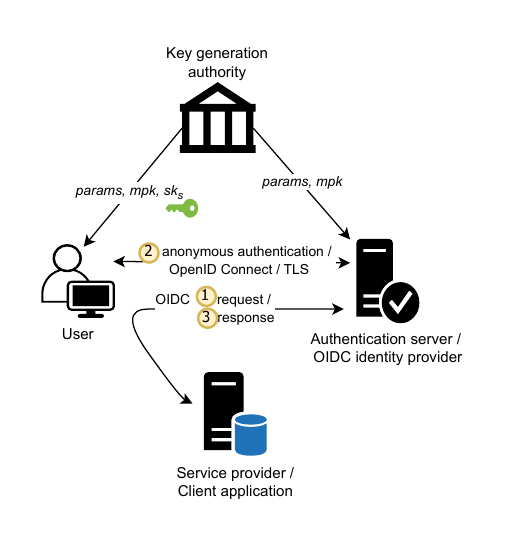}
  \caption{Proposed scenario: a user anonymously authenticates to an authentication server to access a service provided by a service provider.}
  \label{fig:fig1}
\end{figure}

The proposed anonymous authentication scheme enables a user to authenticate to an authentication server to access a resource or service provided by a service provider without revealing its identity. The scheme assumes that the user has already received the ABE secret keys $sk_{S}$ associated with its attributes from a key generation authority. Both the user and the authentication server share the public system parameters $params$. Additionally, the authentication server knows the master public key $mpk$ and the access policy associated with the requested service (the actors of this scenario are illustrated in Figure \ref{fig:fig1}).

The protocol supports both user-initiated and server-initiated flows. For example, the authentication server  may initiate the exchange. Instead of providing key distribution as proposed in the literature (e.g., \cite{TS103532}, \cite{Portnoi2015}), the proposed anonymous authentication scheme enables key agreement between the user and the authentication server. 

\begin{figure}[ht!]
  \centering
\includegraphics[width=0.7\columnwidth]{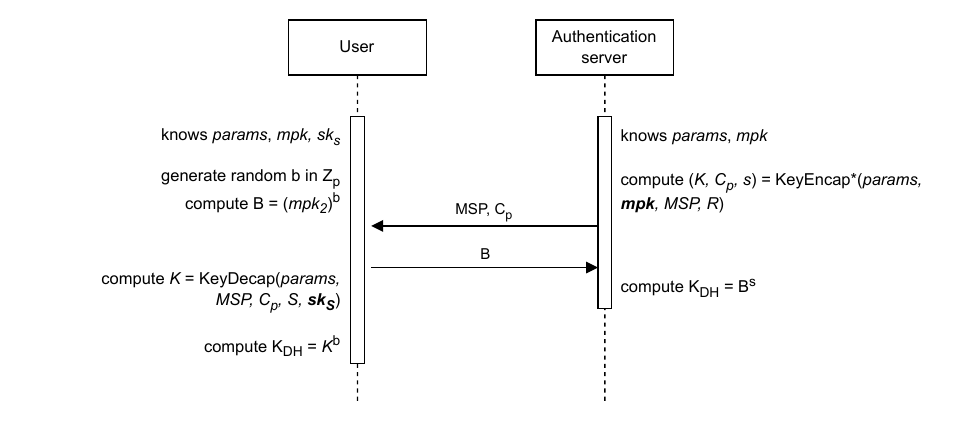}
  \caption{Basic anonymous authentication scheme.}
  \label{fig:fig2}
\end{figure}

As shown in Figure \ref{fig:fig2}, the proposed scheme leverages the Elliptic-Curve Diffie-Hellman Ephemeral (ECDHE) protocol for key agreement, providing forward secrecy through the use of ephemeral session keys. The authentication server encapsulates a Diffie-Hellman partial key $K$ using the ABKEM scheme and retains the corresponding random value $s$. Simultaneously, the user selects a random value $b$ and computes its own partial key $B = g_2^b$. These values are exchanged, allowing both parties to compute the shared Diffie-Hellman key $K_{\mathit{DH}} = K^b = B^s$, which is then used to derive secure session keys.

The ability of the user to communicate through the secure channel is an implicit proof of its ability to decapsulate the ciphertext sent by the authentication server, which demonstrates that it possesses the right set of attributes that satisfy the access policy MSP of the requested service. To explicitly demonstrate the ability of the user to decapsulate the ciphertext, another variant of the proposed scheme involves a key confirmation operation, which enables the scheme to be used for just anonymous authentication i.e., the derived session keys are not used to establish a secure communication channel after authentication but rather as a proof of the ability of the user to decapsulate the ciphertext using the right set of attributes.

\begin{figure}[ht!]
  \centering
\includegraphics[width=0.7\columnwidth]{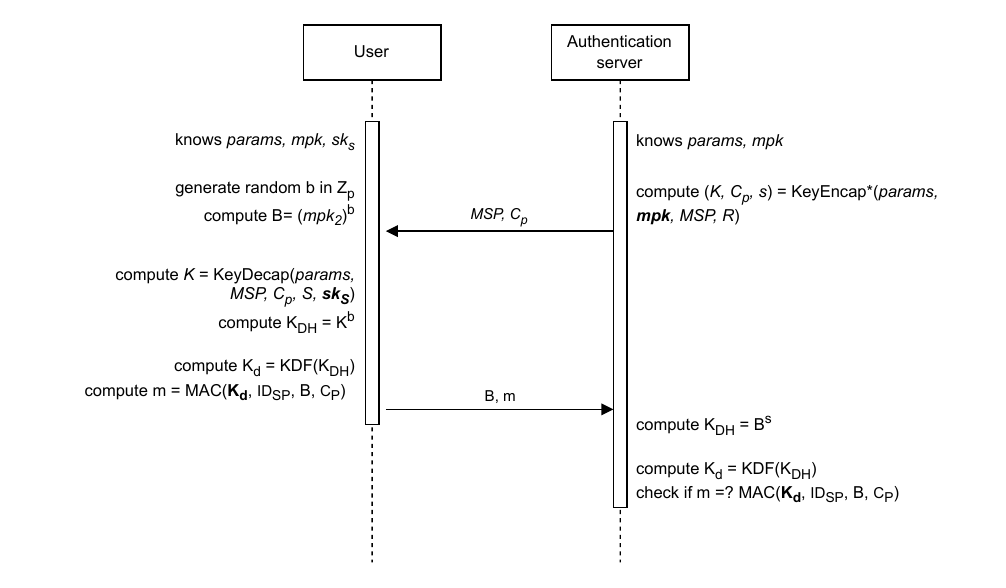}
  \caption{Basic anonymous authentication scheme including a key confirmation operation.}
  \label{fig:fig3}
\end{figure}

In the proposed variant, the user derives from the computed Diffie-Hellman key $K_{DH}$ a session key $K_{d}$ using a Key Derivation Function (KDF) that is based on a pseudorandom function using, for example, a cryptographic hash function or a block cipher. The user then computes a digest $m$ of the public Diffie-Hellman partial keys and the server provider identity $ID_{SP}$, using a Message Authentication Code (MAC) (typically, a keyed cryptographic hash function e.g., HMAC or a block cipher). The computed digest $m$ is transmitted to the authentication server  along with the Diffie-Hellman partial key $B$, as illustrated in Figure \ref{fig:fig3}.

\section{Security and privacy analysis}
This section evaluates whether the proposed anonymous authentication scheme meets the security and privacy requirements defined in Section 4.1.

\subsection{Hardness assumptions}
The security of the proposed anonymous authentication scheme relies on the security of the used \textbf{ABKEM}. For example, the security of the Waters scheme \cite{Waters2011} is based on the hardness of the Bilinear Diffie-Hellman (BDH) problem. The authentication exchange inherits this security assumption, as it uses the same base groups defined in the public system parameters, $params$.

\subsection{Security}
When used in standalone, the proposed anonymous authentication scheme does not guarantee all the security properties defined in Section 4.1. The scheme must be extended with additional interactions between the user and the authentication server  as in the SIGMA protocols family \cite{Krawczyk2003} to provide all the security properties, or coupled with another security protocol or standard like the Transport Layer Security (TLS) protocol. Nonetheless, the scheme does provide authentication for authorised users. 

The variant of the proposed scheme that includes a key confirmation operation ensures authentication on the user side only. To provide secure authentication of the authentication server side i.e., mutual authentication, the proposed simple authentication scheme can be used within the OpenID Connect framework. The framework can provide the authentication of the authentication server using the “server authentication only” mode of the Transport Layer Security (TLS) protocol, which is the most commonly deployed TLS mode in the Internet. The OpenID Connect framework provides, inter alia, a single sign on feature, which allows the user to authenticate once and access several related, yet independent, services.

Thanks to TLS and the key confirmation operation that encloses the server provider identity in the MAC digest, man-in the-middle attacks are mitigated. Moreover, since both parties are generating random values to compute Diffie-Hellman partial keys, these latter serve a dual-use as keys and as nonces. This dual use of Diffie-Hellman partial keys as both keys and nonces ensures resistance to replay attacks.

If the proposed anonymous scheme is used to establish a secure communication channel, perfect forward secrecy is provided owing to the use of ephemeral Diffie-Hellman keys.

Furthermore, the integration of the anonymous authentication scheme with TLS remains compatible with post-quantum versions of TLS, thereby maintaining resilience against cryptographically relevant quantum attacks.

\subsection{Anonymity}
Anonymity is not always a desirable feature, because it is in conflict with the auditability and accountability features that require knowing the users who had access or may have access (before the fact audit) to a resource. Attribute-Based Encryption (ABE) is inherently identity-agnostic. Decryption of the ciphertext is realised on the basis of the attributes the user possesses and presents through the decryption keys; it does not require knowledge of user identity. While this allows for anonymous authentication, it does not guarantee a complete user anonymity in some cases. For instance, if an access policy can be composed such that only few users are in the possession of the attributes that satisfy the access policy, then when the policy is enforced for access control, the user requesting authentication can be guessed with high probability \cite{Lanus2023}. Consequently, it becomes possible to collect user login history and then infer sensitive information associated with the encrypted content. Thus, ABE offers anonymity as a by-product—unless the user identity is explicitly used as an attribute or in special cases. 

In this respect, the proposed scheme offers user anonymity, provided that the following constraints are met:
\begin{itemize}
\item Attributes must not be linked to user identity or Personally Identifiable Information (PII).
\item Access policies corresponding to different user requests must be satisfiable by at least $r$ distinct users to achieve $r$-anonymity, where the probability of user identification is $1/r$. This can be achieved by techniques such as padding, as proposed in \cite{Lanus2023}, which employs anonymising arrays to ensure that any attribute assignment appears at least $r$ times
\end{itemize}

The user needs to verify that the access policy representation MSP exchanged by the authentication server (and associated with the symmetric key encapsulation) satisfies the aforementioned constraints, which demonstrates that the scheme provides verifiable anonymity.

\section{Demonstrator: Anonymous authentication with OpenID Connect}
A variant of the proposed anonymous authentication scheme has been implemented and deployed within an open-source OpenID Connect (OIDC) framework. The OpenID Connect is a framework developed by the OpenID foundation\footnote{https://openid.net/} as a suite of lightweight specifications of an identity layer built on top of the OAuth (Open Authorization) 2.0, which is an authorization protocol standardized by IETF\footnote{https://datatracker.ietf.org/wg/oauth/about/}. As a federated identity technology, the OpenID Connect framework enables client applications to verify the identity of a user based on the authentication performed by an authorisation server, as well as to obtain basic profile information about the user.

Owing to its single-message exchange structure, the proposed scheme can seamlessly replace conventional challenge-response mechanisms (e.g., login-password) within an OIDC implementation.

The anonymous authentication implementation includes a user part and a server part that are both implemented using the Charm toolkit. Charm\footnote{https://github.com/JHUISI/charm}  is a Python framework for rapidly prototyping advanced cryptosystems.

\begin{figure}[ht!]
  \centering
\includegraphics[width=0.7\columnwidth]{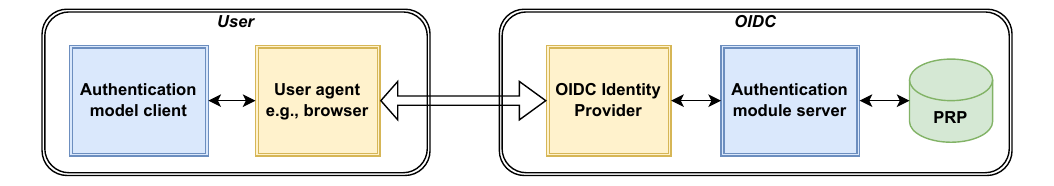}
  \caption{Modules of the implemented system.}
  \label{fig:fig4}
\end{figure}

The server part interacts with a database, emulating a Policy Retrieval Point (PRP), which stores policies associated with client services. A certified OIDC software implementation\footnote{https://openid.net/developers/certified/} was selected and configured to relay authentication requests and responses between the user and server components (see Figure \ref{fig:fig4}). 
 
\begin{figure}[ht!]
  \centering
\includegraphics[width=0.7\columnwidth]{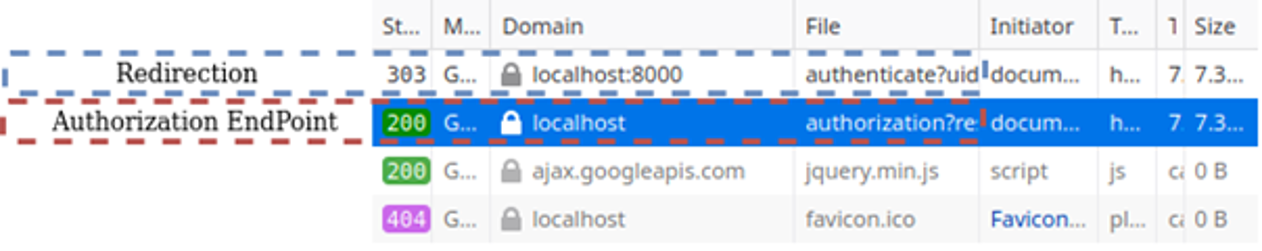}
  \caption{Redirection of authentication request to the user-side ABE module.}
  \label{fig:fig5}
\end{figure}

When requesting to access a service, the user is redirected by the service provider to an authentication server emulating an identity provider with an OIDC authentication request that includes “abe" in the scope. The identity provider calls the ABE-related module. This latter sends a challenge request to the user. The user browser forwards the challenge to the ABE-related module on the user side through a parallel AJAX request. Figure \ref{fig:fig5} shows the redirection at the user side of the authentication request via a POST AJAX request.

\begin{figure}[ht!]
  \centering
\includegraphics[width=0.7\columnwidth]{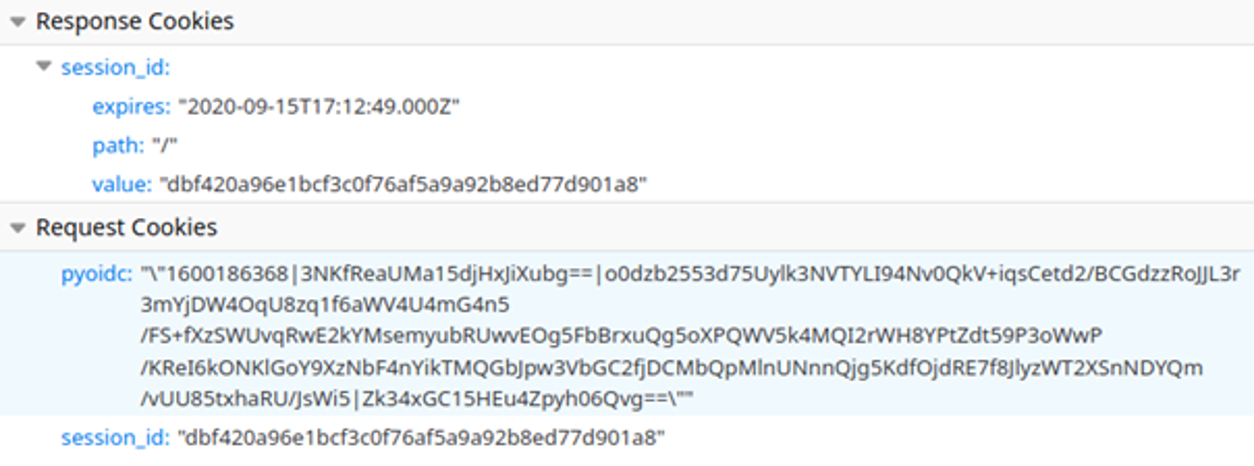}
  \caption{ID token received by the user from the identity provider.}
  \label{fig:fig6}
\end{figure}

The response to the challenge is sent back to the identity provider. After verification of the user response, the identity provider generates an ID token that is sent back within an OIDC response. Figure \ref{fig:fig6} shows the tag “session\_id” value which enables the identity provider to recognise the user end to end through the different redirections. The user records the JWT ID Token as being the value tagged with “pyoidc” which will be decoded, validated and used by the authentication server.

\textbf{Potential application.} The proposed scheme is applicable in a range of domains. For instance, web-based questionnaires are increasingly preferred over postal methods due to their speed and ease of management. There are regarded as an efficient way to provide reporting and collect data about users without jeopardising their anonymity. Thanks to the proposed anonymous authentication scheme, users are able to prove possession of a set of attributes or a predicate that fit a certain policy defining the target questionnaire participants, without revealing their identity.

\section{Remaining challenges}
This section highlights the most important remaining challenges, related mainly to ABE management, which require attention before deploying the proposed solution.

In typical deployments, ABE secret keys are issued by a trusted Key Generation Authority (KGA) and securely distributed to users. KGA is responsible of generating, updating and revoking ABE secret keys, in addition to managing attributes. When a user is compromised or leaves the system, its access rights need to be reduced or revoked by the key generation authority. Updating secret keys or individual attributes is non-trivial, as a change in one attribute can affect a large set of users who share it.

For attribute revocation, an attribute revocation list can be used to indicate the list of attributes that have been revoked, and before any ABE encryption or decryption, all parties check the latest available version of the attribute-revocation list.

A naïve approach to secret key revocation consists of regenerating the entire ABE scheme with redistribution of a new master public key, a new master secret key, and new secret keys \cite{TS103532}. Otherwise, secret key revocation can be handled by introducing time intervals. An expiration timestamp is added to the access policy, and the user needs to use the secret keys that match the expiration condition. Bartolomeo \cite{Bartolomeo2023} proposed another approach using cryptographic accumulators based on bilinear mappings. The revocation of a secret key prevents the decryption of a ciphertext created after the key is revoked, which is called forward revocation. In the proposed approach, the key generation authority assigns an index to each generated secret key, and creates a list of revoked keys. The list, the cryptographic accumulator, and other parameters are made public and updated each time secret keys are revoked. Upon these updates, the users also update locally their unrevoked secret keys. 

The key generation authority that generates all ABE secret keys is given complete power and is implicitly trusted. To alleviate this key escrow problem, it is possible to rely on a multi-authority version of ABE (e.g., \cite{Chase}, \cite{Rouselakis2015}) where there are multiple key generation authorities each of which is responsible for the authorised key distribution of a specific set of attributes.

\section{Conclusion}
In an environment where personal data is increasingly vulnerable, this paper introduced an anonymous authentication scheme that allows individuals to access services securely without compromising their identity. The proposed solution consisted of a single exchange and relied on attribute-based encryption that offers cryptographic enforcement of access control. The scheme enables secure service access while preserving user privacy and offering individuals greater control over their digital footprint.

\section*{Acknowledgement}
The author gratefully acknowledges the contributions of Alaeddine ZAOUALI in the development of the proposed solution. Appreciation is also extended to Université Paris-Saclay and CEA, List for providing the essential resources and facilities that supported this work.

\bibliographystyle{unsrt}  
\bibliography{references}

\end{document}